\documentclass[11pt,twocolumn,prb,superscriptaddress,reprint]{revtex4-1}

\usepackage{graphicx,amsmath,amssymb,color}
\usepackage{tabularx, ctable}

\usepackage{ulem}
\usepackage{bm}
\usepackage{siunitx}
\usepackage{gensymb}

\usepackage[breaklinks=true,colorlinks,allcolors=blue]{hyperref}

\newcommand{\be}{\begin{eqnarray}}
\newcommand{\ee}{\end{eqnarray}}

\begin{document}

\title{Nanopatterning of multicomponent van der Waals heterostructures \\ using atomic force microscopy}

\author{A. L. Shilov$^{+}$}
\affiliation{Department of Materials Science and Engineering, National University of Singapore, 117575 Singapore}
\affiliation{Programmable Functional Materials Lab, Center for Neurophysics and Neuromorphic Technologies, Moscow, 127495}
\affiliation{Moscow Center for Advanced Studies, Kulakova str. 20, Moscow, 123592, Russia}

\author{L. Elesin$^{+}$}
\affiliation{Programmable Functional Materials Lab, Center for Neurophysics and Neuromorphic Technologies, Moscow, 127495}
\affiliation{Institute for Functional Intelligent
Materials, National University of Singapore, Singapore, 117575, Singapore}
\affiliation{Moscow Center for Advanced Studies, Kulakova str. 20, Moscow, 123592, Russia}

\author{A. Grebenko$^{+}$}
\affiliation{Programmable Functional Materials Lab, Center for Neurophysics and Neuromorphic Technologies, Moscow, 127495}

\author{V. I. Kleshch}
\affiliation{Department of Physics, M.V. Lomonosov Moscow State University, Moscow 119991, Russia}

\author{M. A. Kashchenko}
\affiliation{Programmable Functional Materials Lab, Center for Neurophysics and Neuromorphic Technologies, Moscow, 127495}
\affiliation{Moscow Center for Advanced Studies, Kulakova str. 20, Moscow, 123592, Russia}

\author{I. Mazurenko}
\affiliation{Programmable Functional Materials Lab, Center for Neurophysics and Neuromorphic Technologies, Moscow, 127495}
\affiliation{Moscow Center for Advanced Studies, Kulakova str. 20, Moscow, 123592, Russia}

\author{E. Titova}
\affiliation{Programmable Functional Materials Lab, Center for Neurophysics and Neuromorphic Technologies, Moscow, 127495}
\affiliation{Moscow Center for Advanced Studies, Kulakova str. 20, Moscow, 123592, Russia}

\author{E. Zharkova}
\affiliation{Programmable Functional Materials Lab, Center for Neurophysics and Neuromorphic Technologies, Moscow, 127495}
\affiliation{Moscow Center for Advanced Studies, Kulakova str. 20, Moscow, 123592, Russia}

\author{D. S. Yakovlev}
\affiliation{Laboratoire de Physique et d’Etude des Matériaux, ESPCI-Paris, PSL Research University, 75005 Paris, France}

\author{K. S. Novoselov}
\affiliation{Institute for Functional Intelligent
Materials, National University of Singapore, Singapore, 117575, Singapore}

\author{D. A. Ghazaryan}
\affiliation{Laboratory of Advanced Functional Materials, Yerevan State University, Yerevan 0025, Armenia}

\author{V. Dremov}
\affiliation{Programmable Functional Materials Lab, Center for Neurotechnology and Neuromorphic materials, Moscow, 121205, Russia}
\affiliation{Moscow Center for Advanced Studies, Kulakova str. 20, Moscow, 123592, Russia}

\author{D. A. Bandurin$^{*}$}
\affiliation{Department of Materials Science and Engineering, National University of Singapore, 117575 Singapore}

\begin{abstract}
Multilayer van der Waals (vdW) heterostructures have become an important platform in which to study novel fundamental effects emerging at the nanoscale. Standard nanopatterning techniques relying on electron-beam lithography and reactive ion etching, widely applied to pattern such heterostructures, however, impose some limitations on the edge accuracy and resolution, as revealed through numerous experiments with vdW quantum dots and point contacts. Here we present an alternative approach for electrode-free nanopatterning of thick multilayer vdW heterostructures based on atomic force microscopy (AFM). By applying an AC voltage of a relatively small frequency (1-10 kHz) between the sharp platinum tip and the substrate, we realize high-resolution ($\lesssim 100$~nm) etching of thick multicomponent heterostructures if the latter are deposited onto graphite slabs. Importantly, unlike more conventional electrode-free local anodic oxidation, our method does not require a special environment with excess humidity, can be applied at ambient conditions, and enables the patterning of multilayer heterostructures composed of graphene, graphite, hexagonal boron nitride (hBN), NbSe$_{2}$, WSe$_{2}$, and more. 

\begin{center}
$^{+}$ These authors contributed equally. \\
\end{center}

\end{abstract}

\maketitle

The rapid progress in 2D materials research has been enabled by the emergence of unconventional nanofabrication methods. Mechanical exfoliation~\cite{exfoliation}, dry transfer assembly in an inert atmosphere~\cite{high_quality_heter}, encapsulation~\cite{1DContacts, PropertiesOfGrapheneKretinin}, and precise alignment of the crystallographic axes~\cite{MassiveDiracFermions, SmallAngleTBG,CaoMagicAngleTBG,CaoMagicAngleTBG2} are just a few examples of techniques that facilitated a technological leap in creating complex van der Waals (vdW) heterostructures made out of 2D materials which in turn has resulted in the discovery of many spectacular effects. While the patterning of such heterostructures using electron beam lithography (EBL) accompanied by reactive ion etching (RIE), routinely applied to shape such heterostructures in desired forms, has become a gold standard in 2D technology, there are various cases where the homogeneity of the etched edges and the resolution of the obtained structures are not sufficient. Typical examples are quantum dots, nanowires, point contacts, and nanochannels made out of various 2D materials~\cite{DiracBilliard,StaampferQantumDotsInBG, EnsslinQantumDotsInBG,QDWSe2, QDWSe22, InSePC, GeimDeadWater}. To circumvent the  limitations of the etching techniques~\cite{QPCWithRIE,StampferQDEtched,RIEofGNanoribbons} and get a deeper insight into the physical effects occurring in these materials at the nanoscale, alternative patterning methods have been actively developed and applied.

\begin{figure*}[ht!]
  \centering\includegraphics[width=\linewidth]{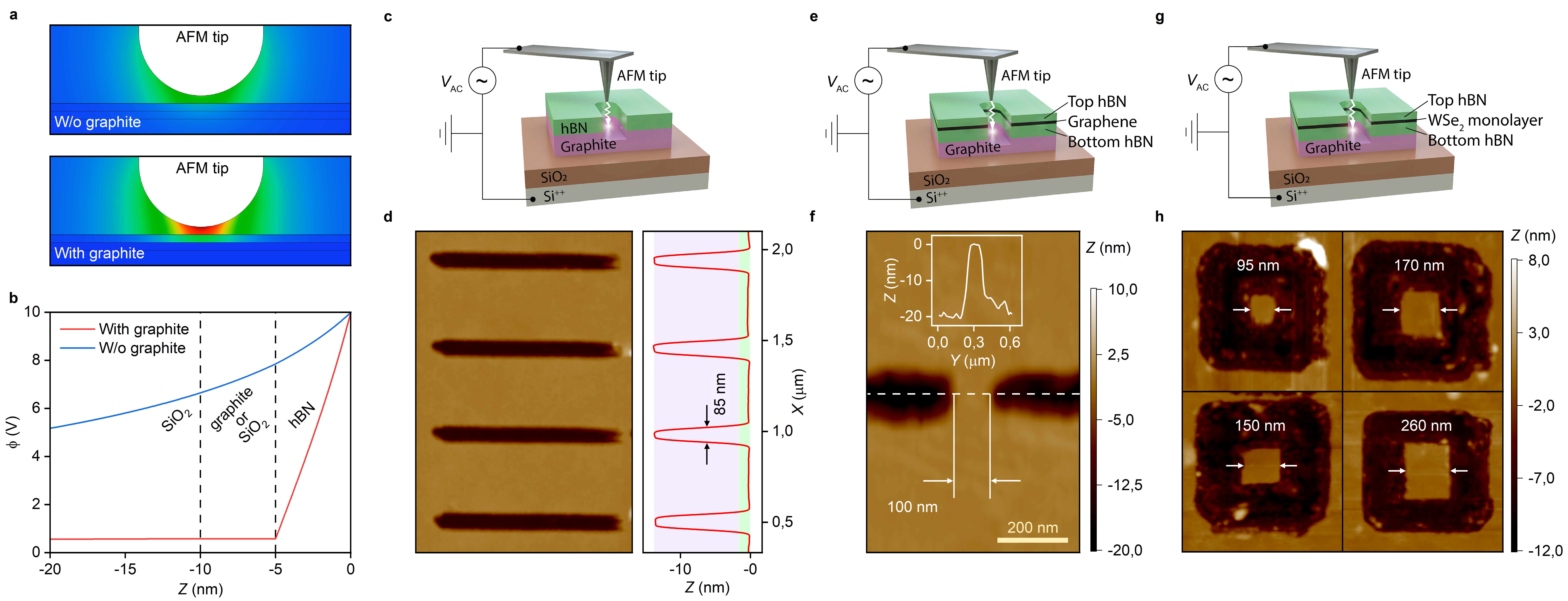}
    \caption{\textbf{AFM nanopatterning of multilayer vdW heterostructures.}
    \textbf{a-b,} Electrostatic simulations of the electric field distribution near the AFM tip, which is positioned in proximity to the heterostructure. The voltage is applied between the tip and the silicon wafer, while the graphite slab remains floating. Figure (a) illustrates the effect of field enhancement due to the graphite base beneath the hBN/graphene heterostructure. The color scale reflects the electric field strength and ranges from 0 V/nm (blue) to 1.7 V/nm (red). Figure (b) shows the $\phi(Z)$  dependences along the axis of symmetry for cases of hBN flake with and without graphite base. \textbf{c-d,} Schematics of the setup and results of AFM nanopatterning in hBN/graphite heterostructure. AFM topography scan features a set of etched trenches with a typical width of 85 nm. \textbf{e-f,} Schematics of the setup and results of AFM nanopatterning in hBN/graphene/hBN/graphite heterostructure. AFM topography scan features a constriction with a width of 100 nm. The inset in (d) demonstrates a defined profile of the constriction, the thickness of both hBN flakes is $\approx$ 3.5 nm. \textbf{g-h,} Schematics of the setup and results of AFM nanopatterning in hBN/WSe$_2$/hBN/graphite heterostructure.  AFM topography scans display a set of nanoscale dots of different sizes.
}
	\label{Fig1}
\end{figure*}

One such technique, scanning probe lithography (SPL), has been recently gaining  momentum. Since the initial experiment on STM~\cite{staufer1987nanometer}, this technique has been improved to enable maskless nanopatterning with the precise alignment of the pattern, relatively high etching rate~\cite{t-SPLofPPAsuperfast} and sub-20 nanometer accuracy~\cite{t-SPLofsilicon,o-SPLofsilicon,eflao}. Compared to conventional EBL accompanied by RIE approach and more arduous focused ion beam fabrication, SPL does not require vacuum operation and prevents samples from contamination with polymer residues or substrate milling products. Several variations of the SPL nanofabrication technique have been introduced over the past decades and have been applied in the processing of a wide range of materials including metals~\cite{o-SPLofgalium,m-SPLofcopper}, semiconductors~\cite{o-SPLofsilicon,o-SPLofsilicon2,o-SPLofMoS2,o-SPLofWSe2,t-SPLofsomeTMDC,m-SPLofGaAs}, graphene or graphite~\cite{o-SPLofgraphene,dobrik2010crystallographically,eflao,m-SPLofgraphene,o-SPLofgraphenenanoribbons,o-SPLofQDongraphene,QPCwithGraphiteGates}, polymer films~\cite{t-SPLofPPAsuperfast,gottlieb2017thermal,xiePolymerSPL2016,t-SPLofPPV,t-SPLofpolycarbonate,t-SPLofsilicon} and organics~\cite{shaw2013demand}. Nonetheless, while techniques such as thermal- and thermochemical-SPL have proven 
fast and accurate~\cite{review_of_tSPL, review}, they require special heated probes, can be only applied to highly thermosensitive materials, and are unsuitable for most solid films. Another promising method, mechanical-SPL~\cite{dong2004nontraditional}, can only be applied  to a limited number of materials and suffers from probe deterioration, debris formation, and surface degeneration~\cite{m-SPLofgraphene,m-SPLofcopper}. The oxidation-SPL and its derivative - electrode-free local anodic oxidation (EFLAO) have been suggested as a versatile alternative for producing structures with defined edges~\cite{ryu2017advanced,eflao}. It has been successful in manufacturing nanoscale structures and has been commonly applied to graphene~\cite{o-SPLofgraphenenanoribbons,o-SPLofgraphene,eflao,QPCwithGraphiteGates}, semiconductors~\cite{o-SPLofMoS2,o-SPLofsilicon,o-SPLofsilicon2}, metals and thin conductive films. However, the oxidation-SPL demanded a high relative humidity (RH$>50\%$) and with only one exception~\cite{o-SPLofgraphenenanoribbons} could only be applied to relatively thin samples. 

In this paper, we propose a simple and accessible method of electrode-free AFM nanopatterning that can be applied to complex multilayer heterostructures composed of hBN, monolayer graphene, few-layer graphite, NbSe$_{2}$, and WSe$_{2}$. The central idea of the method is to exploit the effect of electric field enhancement, attained through the assembly of the hBN-encapsulated vdW heterostructures on conductive graphite flakes - electrostatic simulations in Fig.~\ref{Fig1}a-b show that graphite base can increase the electric field penetrating the structure by an order of magnitude. The strong electric field confined inside the heterostructure initiates the breakdown of the dielectric hBN buffer and launches the chemical reaction that etches the stack. The method features high precision and reproducibility,  allows to perform the lithography in the low humidity ambient (RH $\sim20\%$), and can be applied to heterostructures of practically any complexity. 

\begin{figure*}[ht!]
  \centering\includegraphics[width=0.9\linewidth]{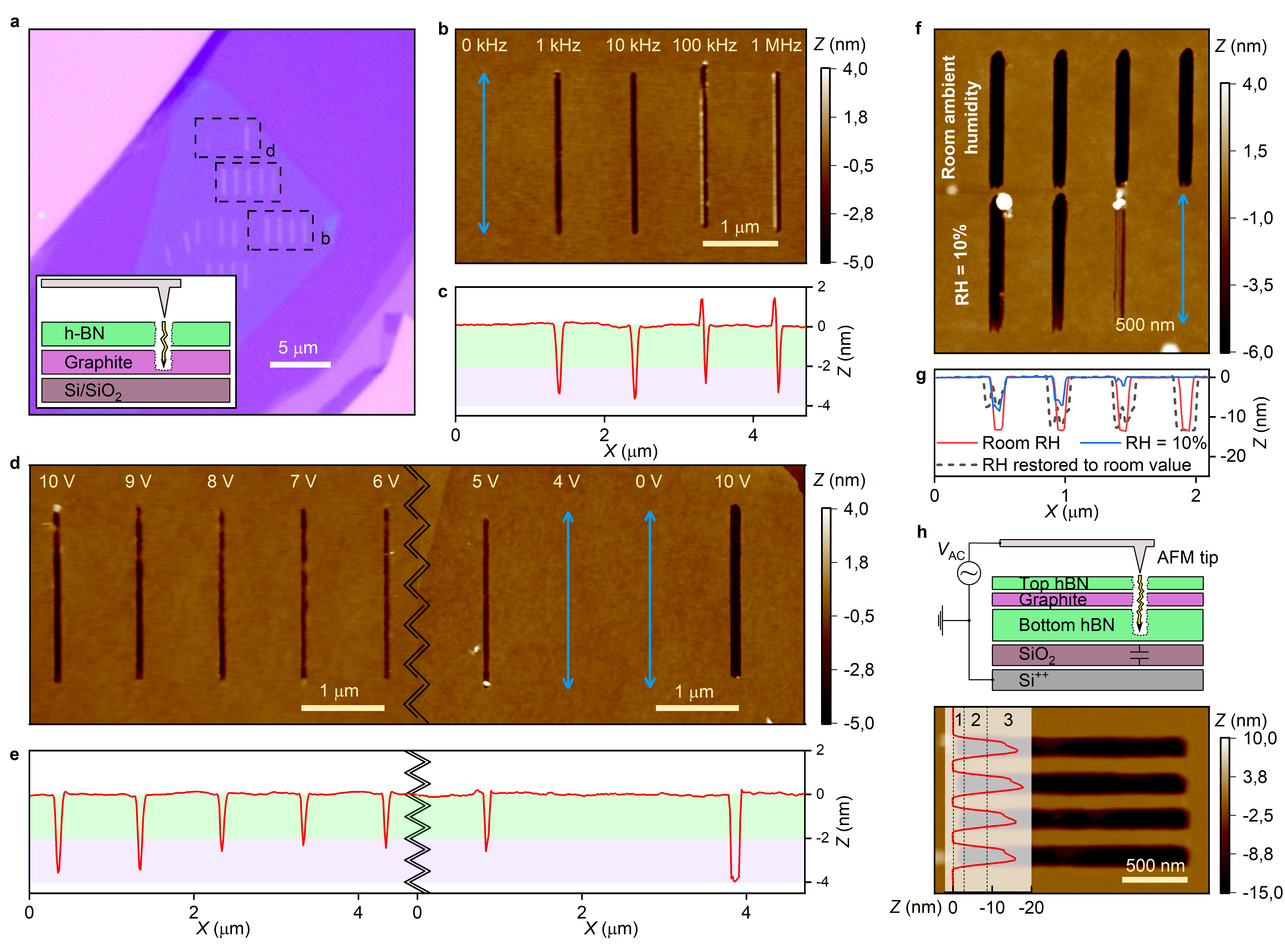}
    \caption{\textbf{Charachterization of the AFM nanopatterning.}
    \textbf{a,} Optical image of the sample displaying a series of engraved patterns, with a schematic of the employed structure included in the inset. RH was kept at 17\%. \textbf{b-c,} Method's operation under different excitation frequencies and at a fixed voltage amplitude of 10~V. The correspondent region is marked in (a). \textbf{d-e,} Method's operation under given bias voltage amplitudes and at a fixed excitation frequency of 10 kHz. The corresponding regions are marked in (a). Profiles shown in (c) and (e) represent the averaged profile of the pattern along the X-axis. Zigzag lines in (d-e) remove pattern-free regions. \textbf{f,} A set of trenches obtained at different RH. The first row of lines was obtained at room RH $\approx 20\%$. The second row was obtained after the RH was reduced to 10\%. Both patterns were acquired using the same AFM probe under 10 kHz excitation with 10 V amplitude. \textbf{g,} Averaged profiles along the X-axis for the sets of trenches, obtained at different RH. \textbf{h,} Schematics of the setup and results of AFM nanopatterning in hBN/graphite/hBN heterostructure. Numbers 1-3 in the inset denote the regions of top hBN, graphite and bottom hBN respectively.}
	\label{Fig2}
\end{figure*}

\textbf{Samples and AFM nanopatterning technique.} To demonstrate the versatility of our method we shot its performance on several types of vdW heterostructures comprised of various 2D materials fabricated using the hot-release dry transfer method described previously~\cite{cleaning_interfaces}. In brief, 2D materials flakes were mechanically exfoliated using a scotch tape method and deposited onto Si/SiO$_2$ substrates. Selected flakes were then sequentially picked up using a polycarbonate (PC) membrane stretched over polydimethylsiloxane (PDMS) polymer block placed on a glass slide attached to the micromanipulator of the home-built transfer station. The resulting heterostructure was released on a clean conducting Si$^{++}$ wafer covered by 285 nm of SiO$_2$ preheated to 200 \celsius. 

Once the heterostructure was fabricated, the Si$^{++}$/SiO$_2$ substrate, on which it was assembled, was attached to the grounded AFM stage (NT-MDT model Next II). A bias voltage with a low frequency of $10$ kHz and an amplitude of $5-10$~V was applied to the AFM tip using the voltage supply circuit integrated into the microscope. Note that the heterostructures remained floating with capacitive coupling between the graphite and the silicon substrate. We used a Pt-coated silicon tip that, prior to the experiment, was annealed in the Ar/H$_2$ atmosphere at 450~\celsius\; for 30~minutes. We also note, that unannealed Pt-coated and W$_2$C-coated silicon probes demonstrated very limited performance: the patterning was poorly reproducible. The patterning was performed by scanning the probe over the sample's surface in the contact regime at 0.5~$\mu$m/s, repeating the same pattern 5-10 times. The RH of the environment was maintained at typical room levels, ranging between $15-20$\%.

Figure~\ref{Fig1}c,e,g provides an overview of the experimental setups we used to pattern various heterostructures. The scope of the operable structures encompasses both basic two-layer stacks, comprised of hBN and graphite (shown in  Fig.~\ref{Fig1}c), and more sophisticated variations such as graphene (or WSe$_2$) monolayers sandwiched between hBN flakes (as shown in Fig.~\ref{Fig1}e,g). We note that despite the apparent compositional distinctions among used structures, they share two common features, namely: all contain graphite slab underneath the heterostructure and are covered by the hBN flake on the top (see below). In Fig.~\ref{Fig1}d,f,h we show the results of using our AFM nanopatterning technique in corresponding heterostructures. The method has been proven capable of producing nanotrenches (Fig.~\ref{Fig1}d), nanoconstrictions (Fig.~\ref{Fig1}f), and nanodots (Fig.~\ref{Fig1}h) with the feature size of about 100 nm. The resolution of our method is supposedly limited only by the sharpness of the AFM tip. The topography of the engraved slits, shown in Fig.~\ref{Fig1}d, features clean surface, high accuracy, and consistency in the dimensions of the obtained pattern. The averaged profile of the pattern reveals high uniformity of the width ($\sim85~$nm) and depth of the lines. Importantly, the line depth was found equal to the total thickness of the sample, pointing out that the structure was fully etched through (Supplementary Information).

\begin{figure*}[ht!]
  \centering\includegraphics[width=0.93\linewidth]{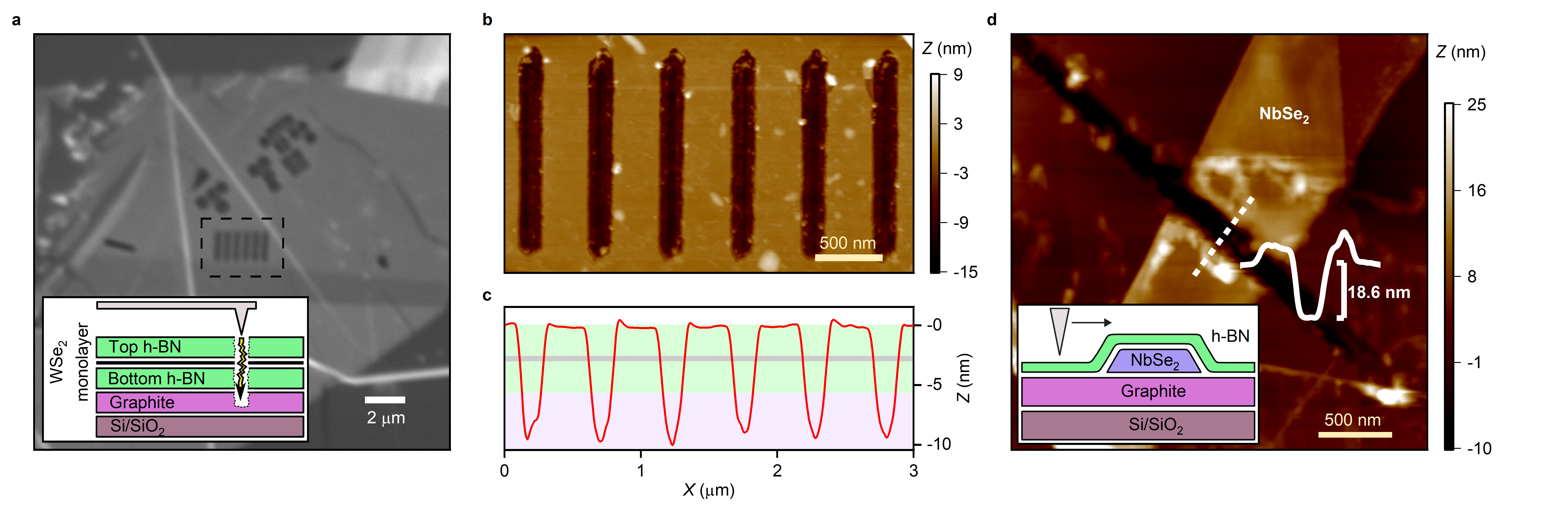}
    \caption{\textbf{AFM nanopatterning of encapsulated TMDC materials.}
    \textbf{a,} Optical image of the heterostructure, containing WSe$_2$ monolayer, which features a set of engraved bands and nanoscale dots. The schematic of the structure is provided in the inset. \textbf{b-c,} AFM topography scan and averaged profile along the X-axis of the region, marked in (a). The voltage of 10 V at 10 kHz was applied; the RH was kept at 17\%. \textbf{d,} NbSe$_2$ flake ($d\sim 10$ nm), sandwiched between hBN and graphite flakes and divided into two parts via AFM nanopatterning technique. The inset shows the schematic of the sample. }
	\label{Fig3}
\end{figure*}

\textbf{Detailed characterization of the AFM nanopatterning.} 
To get further insight into the basic principles underlying our AFM lithography technique and to comprehend its capabilities and limitations we used the simple heterostructure (as presented in Fig.~\ref{Fig1}c), consisting of thin hBN slab (2~nm thick) deposited onto a graphite flake (2~nm thick) (Fig.~\ref{Fig2}a). Figures~\ref{Fig2}b,c display a series of trenches obtained at different excitation frequencies ranging from 0 to 1 MHz with the amplitude of $10~$V. While the heterostructure was found robust with respect to the DC voltage applied between the AFM tip and the grounded stage, already relatively low-frequency excitation of 1kHz resulted in the deep etching of the stack. Although we found that the method is only weakly dependent on the excitation frequency, as the depth of the lines remained consistent throughout the entire range, higher frequency patterning ($0.1-1~$MHz) resulted in mild contamination of the surface close to the slit. We also studied the performance of our patterning technique at different excitation amplitudes and found a threshold-like dependence. The profiles of the engraved trenches, which were created successively with a single tip employed and at a fixed frequency of $10~$kHz, show a consistent correlation between the depth and the voltage amplitude. In particular, the $10~$V excitation resulted in an almost complete cut of the whole stack, the $5~$V excitation  led only to the partial etching of the heterostructure, and no etching was observed when the excitation amplitude was lowered to $4~$V. Importantly, the last line in the series (Fig.~\ref{Fig2}d) was successfully etched at the $10~$V excitation which eliminates the possibility that the negative result at lower voltages was caused by the deterioration of the probe.

The functional range of RH in our AFM-lithography technique is another aspect distinguishing it from conventional local anodic oxidation techniques~\cite{eflao}. Figure~\ref{Fig2}f demonstrates the results of our electrode-free nanopatterning obtained at two different RH levels in another graphite/hBN heterostructure (with a total thickness of 15 nm). The upper row of trenches was obtained in room ambient conditions (RH $\approx 20\%$), while the lower row was patterned using the same AFM probe after the RH was reduced down to 10$\%$ using a pack of silica gel placed in a sealed environment of the AFM enclosure. While the ambient humidity ensured accurate patterning of the whole heterostructure, RH of 10$\%$ was not sufficient to initiate the etching. The reduction of the lithography efficiency, which is associated with the drop in RH, was found to be reversible: upon reverting the RH to the room value (20$\%$), the method resumed its normal operation (Fig.~\ref{Fig2}g), thus excluding possible probe's degradation. This observation also highlights that the role of water cannot be disregarded as it presumably participates in a chemical reaction with hBN yet its excess is not required.

For further exploration of the processes governing the etching, we released a graphite/hBN heterostructure on a thick hBN flake (see Fig.~\ref{Fig2}h). While it was initially anticipated that only the top two layers would be affected by the AFM nanopatterning, the etching process affected the lower hBN flake as well. The trenches in the bottom hBN flake go down by 8 nm under the graphite. This hints at the presence of an intense chemical reaction taking place during the patterning (see Fig.~\ref{Fig2}h).

\textbf{Nanopatterning of other encapsulated 2D materials. } Following the characterization of the AFM-lithography on basic heterostructures, we investigated other materials that can be patterned using this method. To this end, we used WSe$_2$, for which we produced a heterostructure consisting of an hBN-encapsulated WSe$_2$ monolayer deposited on a graphite slab. Figures~\ref{Fig3}a,b shows the surface topography of the sample after the AFM nanopatterning was performed, showing an array of etched trenches of $150~$nm in width. In the same sample, we also designed a set of nanoscale dots (see Figure~\ref{Fig1}h), confirming the method's efficacy in producing highly accurate submicron patterns. Another interesting result was observed in heterostructures, which contained thick ($\sim$ 10 nm) flakes of 2D superconductors: NbSe$_2$ and Bi$_2$Sr$_2$CaCu$_2$O$_{8+\delta}$ (see. Fig. \ref{Fig3}d and Supplementary Information). Once launched in the vicinity of the flake where only the graphite underlies hBN, the etching process persisted within the NbSe$_2$ flake's specific region, dividing it into two distinct parts. Strikingly, if the etching started directly in the region of the NbSe$_2$ flake, it did not result in a noticeable surface modification. This observation indicates the important role of hBN/graphite combination.

\textbf{Disentangling the origin of AFM nanopatterning.} While the experimental settings we used to pattern various multilayer heterostructures resemble those used in local anodic oxidation nanopatterning studies, there is a number of discrepancies undermining EFLAO as the underlying physical mechanism in our experiments.  Local anodic oxidation involves the formation of a water bridge between the tip and the surface of the sample, which serves as an electrochemical cell where oxygen-containing radicals are generated~\cite{Mechanism_of_o-SPL}. These radicals interact with the sample, leading to surface modification. Supporting this model, previous research demonstrates a strong correlation between etching performance and the RH of the environment~\cite{eflao}. Furthermore, previous studies indicate that the electrochemical reaction is regulated by the voltage drop across the water bridge, which is highly sensitive to the frequency of the applied voltage due to the capacitive coupling between the sample and the silicon substrate. 

These considerations somewhat contradict the observations presented in our work: we demonstrated efficient nanopatterning over a broader range of RH and excitation frequencies. Moreover, previous studies revealed a high resistivity of hBN flakes to oxidation~\cite{hBN_resistant_to_oxydation}, which casts doubt on the assumption that the oxidation of hBN is the sole or major mechanism responsible for the etching process. This conjecture is corroborated by the observation that even thin hBN flakes deposited onto Si$^{++}$/SiO$_2$ substrate not covered by graphite could not be patterned regardless of their thickness, the RH of the environment, and excitation frequency or amplitude. Additionally, the very premise of an electrochemical reaction present in a water bridge between a tip and a sample becomes irrelevant because most of the voltage drop builds up across the highly resistant hBN flake (Fig. ~\ref{Fig1}b).

While it is apparent that the etching of hBN cannot be attributed exclusively to the electrochemical reactions, the mechanism of electrical breakdown can potentially play the predominant role. When the electric field inside the hBN flake beneath the tip apex exceeds the breakdown field $E_{\text{BD}}$ of hBN, it results in a giant current surge that may trigger an intense chemical reaction involving hBN, graphite, and H$_2$O. The products of the reaction can be CO$_2$ from the oxidation of graphite and NH$_3$ together with the boric acid H$_3$BO$_3$ as a consequence of hBN hydrolysis~\cite{hBN-hydrolysis} that in turn may etch encapsulated 2D material. Since the formation of a water bridge is no longer necessary, the method is able to operate at much lower humidity. Voltage frequency also becomes insignificant as the breakdown is governed by the amplitude. According to the previous studies~\cite{anisotr_hbn_breakdown,hbn_breakdown}, the typical value of $E_{\text{BD}}$ is 1-2 V/nm for thin hBN flakes, which limits the thickness of operable hBN flakes to $\sim10$  nm at 10~V bias voltage amplitude. This is also consistent with the threshold value of $\approx 4$ V observed for a 2-nm hBN flake in our study (see Fig. \ref{Fig2}d,e).

In conclusion, we have demonstrated electrode-free nanopatterning of thick multicomponent vdW heterostructures by applying low-frequency voltage between the AFM tip and the grounded AFM stage on which the heterostructures were deposited. The key components of the method are 1) graphite slabs on which the stacks were assembled, 2) hBN layer that covers the whole heterostructure. Conductive graphite slabs localize and amplify the electric field from the tip on the heterostructure initiating the nanopatterning through the chemical reaction launched by the hBN breakdown. While the exact mechanism underlying the etching process still remains to be uncovered, our AFM nanopatterning can be applied for high-resolution shaping of complex vdW heterostructures comprising graphene, hBN, graphite, WSe$_2$, NbSe$_2$, Bi$_2$Sr$_2$CaCu$_2$O$_{8+\delta}$, and potentially other materials too.

\vspace{1em}

\noindent\rule{6cm}{0.4pt}

*Correspondence to: dab@nus.edu.sg




\section*{Competing interests}
The authors declare no competing interests.


%

\newpage
\setcounter{figure}{0}
\renewcommand{\thesection}{}
\renewcommand{\thesubsection}{S\arabic{subsection}}
\renewcommand{\theequation} {S\arabic{equation}}
\renewcommand{\thefigure} {S\arabic{figure}}
\renewcommand{\thetable} {S\arabic{table}}

\end{document}